\documentclass[twocolumn]{aastex63}
\usepackage{graphicx}
\usepackage{amsmath}
\usepackage{apjfonts,natbib}
\usepackage{appendix}

\usepackage{amsmath}

\newcommand{\gsim}{\lower.5ex\hbox{$\; \buildrel > \over \sim \;$}}
\newcommand{\lsim}{\lower.5ex\hbox{$\; \buildrel < \over \sim \;$}}

\newcommand{\civ}{\hbox{C\,{\sc iv}}}
\newcommand{\siiv}{\hbox{Si\,{\sc iv}}}

\newcommand{\kms}{\ifmmode {\rm km\,s}^{-1} \else km\,s$^{-1}$ \fi}

\newcommand{\ergcms}{\ifmmode {\rm ergs\,cm}^{-2}\,{\rm s}^{-1} \else ergs\,cm$^{-2}$\,s$^{-1}$\fi}
\newcommand{\ergcmsA}{\ifmmode{\rm ergs}\, {\rm cm}^{-2}\,{\rm s}^{-1}\,{\rm\AA}^{-1} \else ergs\, cm$^{-2}$\, s$^{-1}$\, \AA$^{-1}$\fi}
\newcommand{\ergcmsHz}{\ifmmode{\rm ergs\,cm}^{-2}\,{\rm s}^{-1}\,{\rm Hz}^{-1} \else ergs\,cm$^{-2}$\,s$^{-1}$\,Hz$^{-1}$\fi}
\newcommand{\phcms}{\ifmmode {\rm ph\,cm}^{-2}\,{\rm s}^{-1} \else ,ph\,cm$^{-2}$\,s$^{-1}$\fi}
\newcommand{\phcmsA}{\ifmmode {\rm ph\,cm}^{-2}\,{\rm s}^{-1}\,{\rm\AA}^{-1} \else ph\,cm$^{-2}$\,s$^{-1}$\,\AA$^{-1}$\fi}

\newcommand\Msun{\ifmmode M_{\odot} \else $M_{\odot}$\fi}
\newcommand\msun{\ifmmode M_{\odot} \else $M_{\odot}$\fi}
\newcommand\Lsun{\ifmmode L_{\odot} \else $L_{\odot}$\fi}
\newcommand\Zsun{\ifmmode Z_{\odot} \else $Z_{\odot}$\fi}
\newcommand\mpyr{\ifmmode \Msun\,{\rm yr}^{-1} \else $\Msun\,{\rm yr}^{-1}$ \fi}

\newcommand{\Luv}{\ifmmode L_{1450} \else $L_{1450}$\fi}
\newcommand{\Lop}{\ifmmode L_{5100} \else $L_{5100}$\fi}
\newcommand{\Lthree}{\ifmmode L_{3000} \else $L_{3000}$\fi}
\newcommand{\lledd}{\ifmmode L/L_{\rm Edd} \else $L/L_{\rm Edd}$\fi}
\newcommand{\ledd}{\ifmmode L_{\rm Edd} \else $L_{\rm Edd}$\fi}
\newcommand{\lamLlam}{\ifmmode \lambda L_{\lambda} \else $\lambda L_{\lambda}$\fi}
\newcommand{\lbol} {\ifmmode L_{\rm bol} \else $L_{\rm bol}$\fi}
\newcommand{\llbol}{\ifmmode \log\left(\lbol/\ergs\right) \else $\log\left(\lbol/\ergs\right)$\fi}

\newcommand{\fuv}{\ifmmode f_{\lambda}\left(1450\AA\right) \else $f_{\lambda}\left(1450 {\rm \AA}\right)$\fi}
\newcommand{\fthree}{\ifmmode f_{\lambda}\left(3000\AA\right) \else $f_{\lambda}\left(3000{\rm \AA}\right)$\fi}
\newcommand{\fH}{\ifmmode f_{\lambda}\left(1.65\micron\right) \else
$f_{\lambda}\left(1.65\micron\right)$\fi}

\newcommand{\mbh}{\ifmmode M_{\rm BH} \else $M_{\rm BH}$\fi}
\newcommand{\lmbh}{\ifmmode \log\left(\mbh/\Msun\right) \else $\log\left(\mbh/\Msun\right)$\fi}

\begin{document}


\title{A sharp rise in the detection rate of broad absorption line variations in a quasar SDSS J141955.26+522741.1}


\correspondingauthor{Zhicheng He, Guilin Liu}
\email{zcho@ustc.edu.cn, glliu@ustc.edu.cn}
\author{Qinyuan Zhao}
	\affiliation{Key laboratory for Research in Galaxies and Cosmology,
Department of Astronomy, University of Science and Technology of China,
Chinese Academy of Sciences, Hefei, Anhui 230026, China}
   \affiliation{School of Astronomy and Space Sciences,
University of Science and Technology of China, Hefei, Anhui 230026, China}

\author[0000-0003-3667-1060]{Zhicheng He}
	\affiliation{Key laboratory for Research in Galaxies and Cosmology,
Department of Astronomy, University of Science and Technology of China,
Chinese Academy of Sciences, Hefei, Anhui 230026, China}
   \affiliation{School of Astronomy and Space Sciences,
University of Science and Technology of China, Hefei, Anhui 230026, China}

\author{Guilin Liu }
	\affiliation{Key laboratory for Research in Galaxies and Cosmology,
Department of Astronomy, University of Science and Technology of China,
Chinese Academy of Sciences, Hefei, Anhui 230026, China}
   \affiliation{School of Astronomy and Space Sciences,
University of Science and Technology of China, Hefei, Anhui 230026, China}

\author{Tinggui Wang}
	\affiliation{Key laboratory for Research in Galaxies and Cosmology,
Department of Astronomy, University of Science and Technology of China,
Chinese Academy of Sciences, Hefei, Anhui 230026, China}
   \affiliation{School of Astronomy and Space Sciences,
University of Science and Technology of China, Hefei, Anhui 230026, China}

\author{Hengxiao Guo}
	\affiliation{Department of Physics and Astronomy, 4129 Frederick Reines Hall, University of California, Irvine, CA, 92697-4575, USA}
	
\author{Lu Shen}
	\affiliation{Key laboratory for Research in Galaxies and Cosmology,
Department of Astronomy, University of Science and Technology of China,
Chinese Academy of Sciences, Hefei, Anhui 230026, China}
   \affiliation{School of Astronomy and Space Sciences,
University of Science and Technology of China, Hefei, Anhui 230026, China}

\author{Guobin Mou}
   \affiliation{School of Physics and Technology, Wuhan University, Wuhan 430072, China}
\begin{abstract}
We present an analysis of the variability of broad absorption lines (BALs) in a quasar SDSS J141955.26+522741.1 at $z=2.145$ with 72
observations from the Sloan Digital Sky Survey Data Release 16 (SDSS DR16). 
The strong correlation between the equivalent widths of BAL and the continuum luminosity, reveals
that the variation of BAL trough is dominated by the photoionization.
The photoionization model predicts that when the time interval $\Delta T$ between two observations is longer than 
the recombination timescale $t_{\rm rec}$, the BAL variations can be detected. 
This can be characterized as a "sharp rise" in the detection rate of BAL variation at $\Delta T=t_{\rm rec}$.
For the first time, we detect such a "sharp rise" signature in the detection rate of BAL variations.
As a result, we propose that the $t_{\rm rec}$ can be obtained from the "sharp rise" of the detection rate of BAL variation.
It is worth mentioning that the BAL variations are detected at the time-intervals less than the $t_{\rm rec}$ for half 
an order of magnitude in two individual troughs. This result indicates that there may be multiple components with different $t_{\rm rec}$ 
but the same velocity in an individual trough.
\end{abstract}



\section{Introduction} 
\label{sec:introduction}

Quasar outflows, as an essential component of the quasar structure, have been considered playing an important role in the co-evolution of the 
central supermassive black holes (SMBHs) and their host galaxies \citep{Silk1998,loeb2004,Springel2005,novak2011,Soker2011,Choi2014,nims2015,ciotti2017}.
In 10–40\% of the quasars, the central source and outflowing gas aligned on the line of sight, outflows may manifest themselves as broad absorption lines 
(BALs) \citep{Weymann1991,Reichard2003,arav2008,Knigge2008,Scaringi2009,Allen2011}.
Technically, they are defined as BAL/mini-BALs troughs with velocity widths $2000\ \kms$ ($500-2000\ \kms$ for mini-BALs) and at depths $>10\% $ below 
the continuum \citep{Weymann1991,hamann2004}.

It is well known that the BAL troughs are able to vary over timescales from days to years 
\citep{Capellupo2011,Capellupo2012,FilizAk2012,arav2013,FilizAk2013,he2014, he2015,he2017,grier2015,hemler2019,zhang2015, shi2017,sun2017,lu2019}. 
Two main mechanisms lead to BAL variations: (1) changes in the ionization of gas; and (2) absorbing gas moving in and out of the line of sight.
In either case, the BAL variation can provide important clues regarding the origins and the physical conditions of outflows.
For case (1), the variability timescale can constrain the recombination timescale of the absorbing gas, which in turn deduces the gas density
\citep{Barlow1992,wang2015,he2019}. \cite{he2019} proposed that the fraction curve (i.e., the detection rate, see Section \ref{sec:fraction} for details) 
of BAL variation is the integral function of the distribution of recombination timescales. 
Following this method, for an individual BAL outflow, the fraction curve would be a quasi step function. 
The recombination timescale would correspond to the time interval at the "sharp rise" of the fraction curve. 

In this letter, we report the first detection of such a "sharp rise" phenomenon of the fraction curve in a BAL quasar 
SDSS 141955.26+522741.1 at $z=2.145$ from the SDSS DR16. 
This object has 72 observations with the  signal-to-noise ratio (S/N) level at $g$ band greater than 5 in all the epochs. 
This object was found to have a strong correlation between the BAL trough and the continuum flux by previous studies (e.g., \citealt{hemler2019,lu2019}).
The strong correlation and multiple observations make this object an ideal laboratory for the BAL studies.
In Section \ref{sec:evidence}, we present the evidence for photoionization-driven BAL variations. In Section \ref{sec:fraction} and \ref{sec:obsbal}, we 
measure the fraction curve of BAL variations. The conclusions are in Section \ref{sec:conclusions}.
Throughout this work, we adopt a standard $\Lambda$CDM cosmology with with $H_{0} =70$ km~s$^{-1}$~Mpc$^{-1}$, $\Omega_{m} = 
0.3$, and $\Omega_{\Lambda} = 0.7$. 

\section{Evidence for photoionization-driven BAL variations} \label{sec:evidence}
\subsection{The spectral fitting}  \label{sec:fitting}

\begin{figure*}[htb]
\centering
\includegraphics[width=0.4\textwidth]{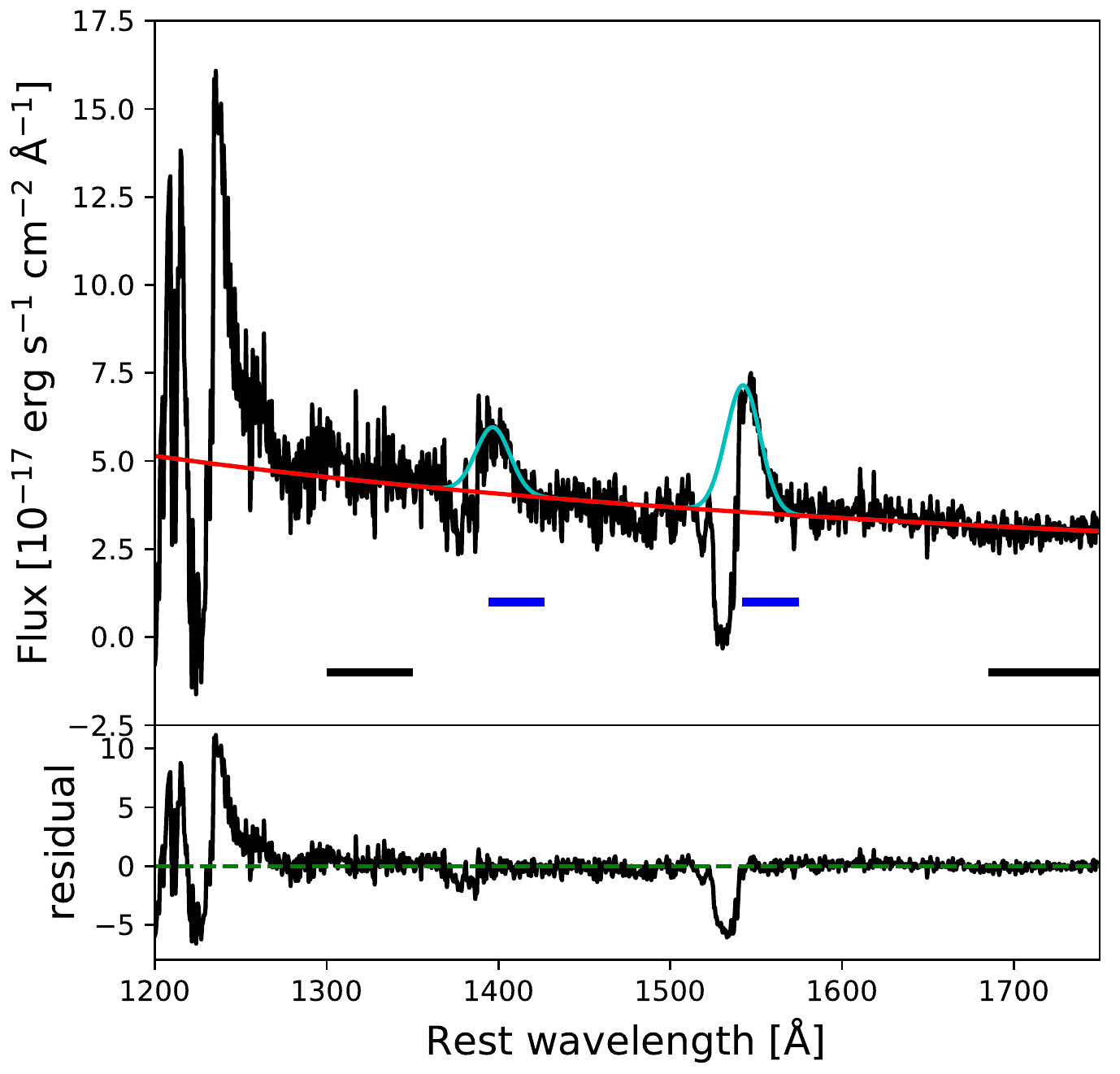}
\includegraphics[width=0.5\textwidth]{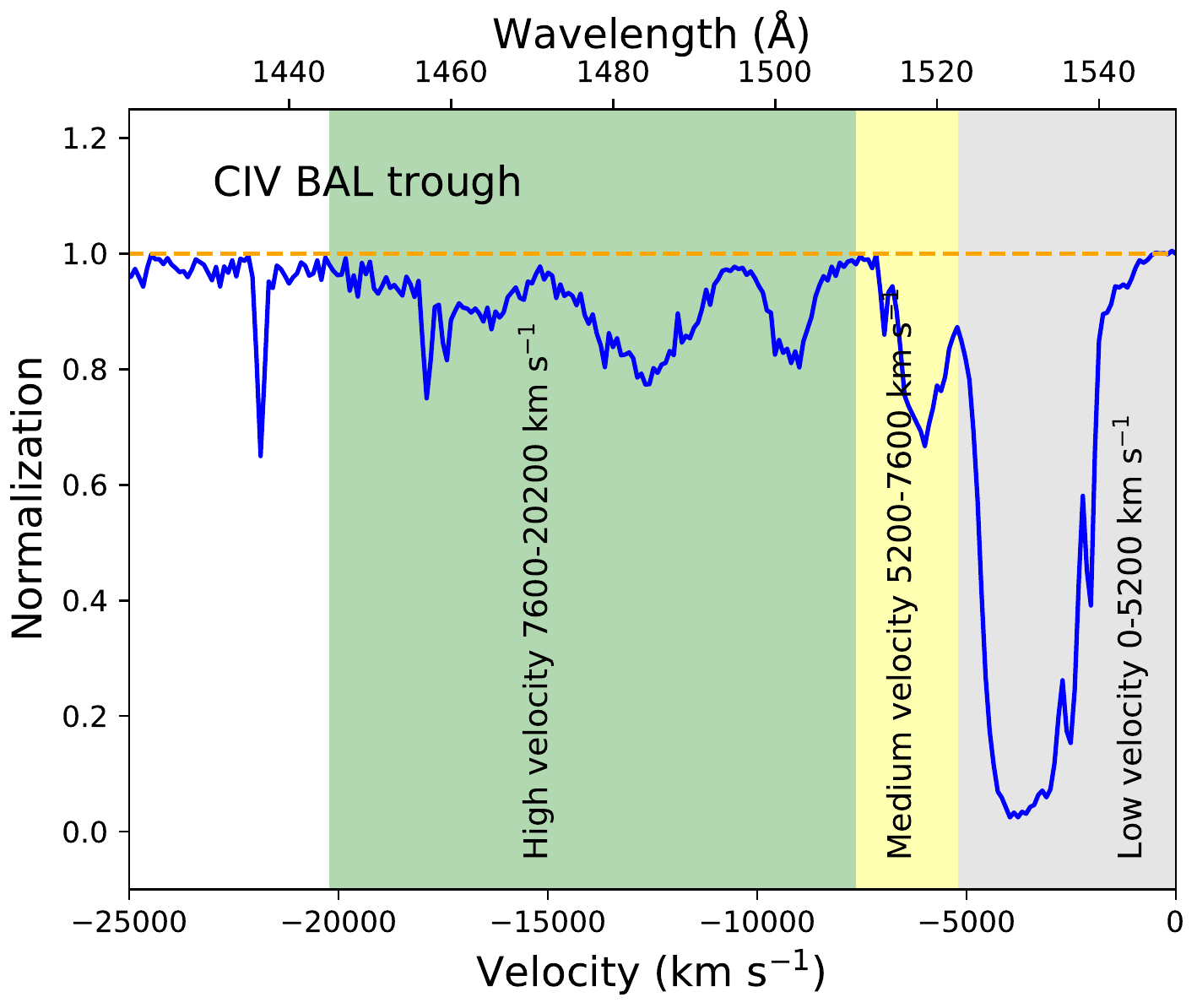}
\caption{Left panel: An example of the power-law plus Gaussian functions to fit the continuum and emission lines. The black line represents the rest-frame spectrum. 
The red line is the power-law continuum fitting and the cyan lines are the Gaussian fitting for the \siiv\ and \civ\ emission lines. 
The black and blue horizontal lines are the fitting windows for the continuum and emission lines, respectively.
The residual spectrum, i.e., the difference between the observational and model spectrum.
Right panel: The normalized composite spectrum for the \civ\ BAL trough from the 72 observations. We divide the whole trough into
three regions: low- (0-5200 \kms, gray), medium- (5200-7600 \kms, yellow) and high-velocity (7600-20200 \kms, green). 
Due to the low detection rate of BAL variations, the three high-velocity regions are treated as one region.
\label{fig1}}
\end{figure*}

We adopt the power-law function $f_{\lambda}= f_{2000} (\lambda/2000)^{\alpha}$ to fit the continuum of the spectra in the `continuum' windows (shown in Figure \ref{fig1}) 
which are known to be relatively free from strong emission lines. 
We use a signal Gaussian profile to fit the \siiv\ and \civ\ emission lines in the continuum-subtracted spectra. 
The absorption regions are masked by the visual inspection when fitting emission lines. As shown in Figure \ref{fig1}, a single Gaussian can well fit the emission lines.

The equivalent widths (EW) of the BAL troughs are calculated as follows:
\begin{equation}
EW=\int {\rm C} [1-\frac{f_{\rm obs}(\lambda)}{f_{\rm con}(\lambda)}]\rm d\lambda,
\end{equation}
where $f_{\rm obs}$ and $f_{\rm con}$ are the observed flux 
and power-law continuum flux, respectively and $\rm C=1$ at $f_{\rm obs} <f_{\rm con}$, otherwise $\rm C=0$.
According to the normalized composite spectrum (right panel of Figure \ref{fig1}), the BAL trough consists of five components at different 
velocities \citep{hemler2019}. We divide the BAL trough into three regions: low- (0-5200 \kms), medium- (5200-7600 \kms) 
and high-velocity (7600-20200 \kms). Note that, due to the low detection rate of BAL variations, the three high-velocity regions are treated as one region.
The EWs are integrated from 1523\AA\ to 1550\AA, 1510\AA\ to 1523\AA\ and 1445\AA\ to 1510\AA\ for the three regions, respectively.
The error for the BAL EW is measured as follows:
$\sigma_{EW}=\sqrt{\sum (\frac{f_{\rm obs}}{f_{\rm con}})^2[(\frac{\sigma_{f_{\rm obs}}}{f_{\rm obs}})^2+(\frac{\sigma_{f_{\rm con}}}{f_{\rm con}})^2]}$,
where $\sigma_{f_{\rm obs}}$ and $\sigma_{f_{\rm con}}$ are the errors in the observed flux 
and power-law continuum flux, respectively. We take the luminosity of the fitted power-law at 1500\AA\ ($L_{1500}$) as the representation of the continuum intensity.

\subsection{The strong negative correlation between the BAL EW and the continuum}  \label{sec:anti}
\begin{figure}[htb]
\centering
\includegraphics[width=0.45\textwidth]{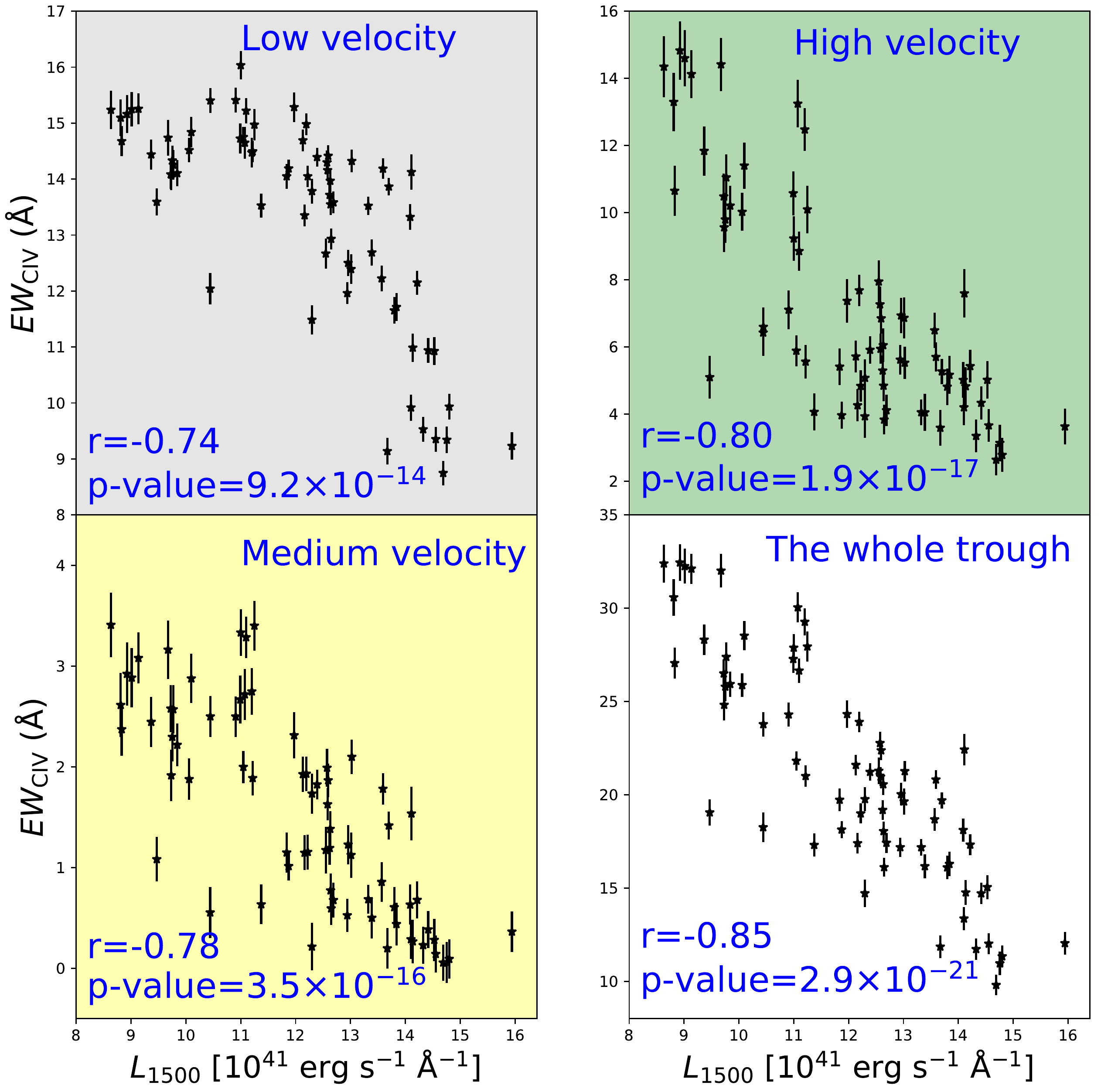}
\caption{The BAL EW of \civ\ versus the luminosity at 1500\AA\ for three velocity regions as well as the whole BAL.
The correlation coefficients and p-values are marked in each panel.}

\label{fig2}
\end{figure}

The change in the ionization state of a outflow gas can be caused by the variation of the incident ionizing continuum.
For a gas in the low ionized state of a specified ion (such \civ), there will be a positive correlation between the \civ\ column density and the continuum luminosity.
In the opposite case, there will be a negative correlation.
Same as reported by \cite{hemler2019} and \cite{lu2019}, there are strong negative correlations between the BAL EW and the continuum luminosity 
at 1500\AA\ for all the three regions (see Figure \ref{fig2}).
The correlation coefficients and p-values are marked in Figure \ref{fig2}.
The strong negative correlations between the BAL troughs and the continuum reveal that the gas is in the overionized state of \civ, and the 
variation of BAL trough is driven by the variation of ionizing continuum. 
As a result, this object is an ideal laboratory to analyze the photoionization-driven BAL variations.

\section{The fraction curve of BAL variations in an individual source}  \label{sec:fraction}
As described in \cite{he2019}, the ionization state of a gaseous outflow requires a period of time 
(the recombination timescale, $t_{\rm rec}$, \citealt{Barlow1992,Krolik1995,wang2015}) 
to respond to changes in the ionizing continuum for the ionized outflows. 
The gas ionization is connected to the average intensity of the ionizing continuum over $t_{\rm rec}$.
We denote the probability of detecting the variability of a BAL with $t_{\rm rec}$ at $\Delta T$ as $p(t_{\rm rec}, \Delta T)$.
In principle, variability of absorption line line can be detected, only when the time interval ($\Delta T$) between two observations longer than 
the recombination timescale, i.e.,  $p=0$ for $ \Delta T <t_{\rm rec}$ and $p=\rm K $ (K is a constant greater than 0) for $\Delta T \ge t_{\rm rec}$.
We define the number ratio of the pair of observations with varied BAL
to all the pairs of observations as the $F(\Delta T)$.
As shown in Eq. 1 in \cite{he2019}, the fraction $F(\Delta T)$ can be written as the integral function
of the distribution of recombination timescale $f(t_{\rm rec})$:
\begin{equation}
\begin{split}
 F(\Delta T) &= \int^{+\infty}_{0} p(t_{\rm rec},\Delta T)f(t_{\rm rec}) dt_{\rm rec}  \\
&= {\rm K} \int^{\Delta T}_0f(t_{\rm rec}) dt_{\rm rec}.
\end{split}
\end{equation}

For an individual source, the $t_{\rm rec}$ is a single value. 
The distribution of recombination timescale $f(t)$ is a $\delta$ function: $f(t)=0\ (t \neq t_{\rm rec})$ and $\int^{+\infty}_{-\infty} f(t) dt=1 $.
As a result, the $F(\Delta T)$ is a step function:

\begin{equation}
F(\Delta T)=\left\{
\begin{aligned}
0~~~~~~~~~~~~~~  &,&\      \Delta T < t_{\rm rec} \\
{\rm K}\ ({\rm K}>0)&,&\         \Delta T \geq t_{\rm rec}.
\end{aligned}
\right.
\end{equation}
In the actual observations, the $F(\Delta T)$ is a quasi step function and the measured $t_{\rm rec}$ is a Gaussian function with a certain width.
So, it is expected that there will be a "sharp rise" phenomenon in the measured $F(\Delta T)$ curve around $ \Delta T =t_{\rm rec}$. 

\section{The observed fraction curve of BAL variations} \label{sec:obsbal}

\begin{figure*}[htb]
\centering
\includegraphics[width=0.45\textwidth]{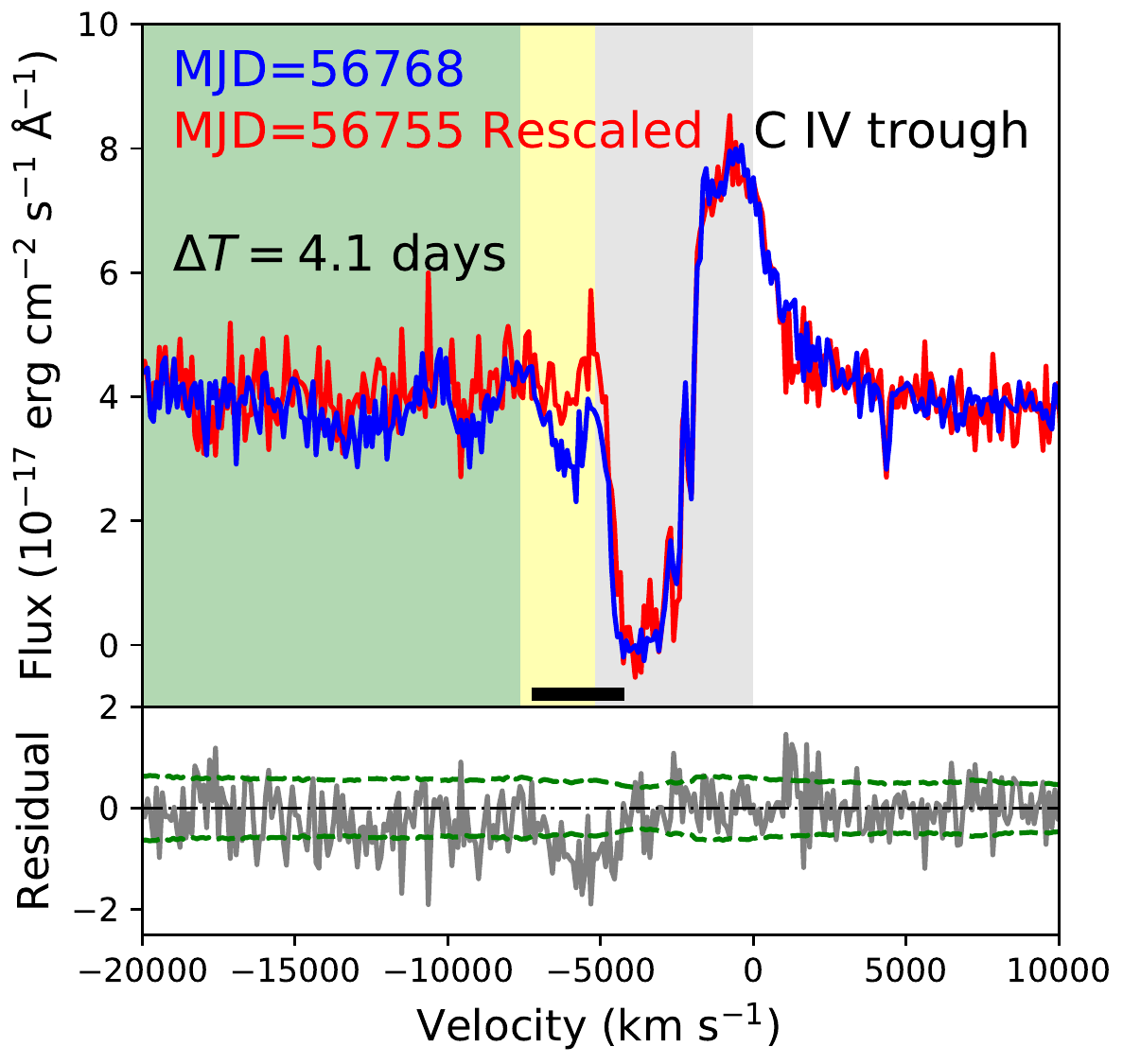}
\includegraphics[width=0.45\textwidth]{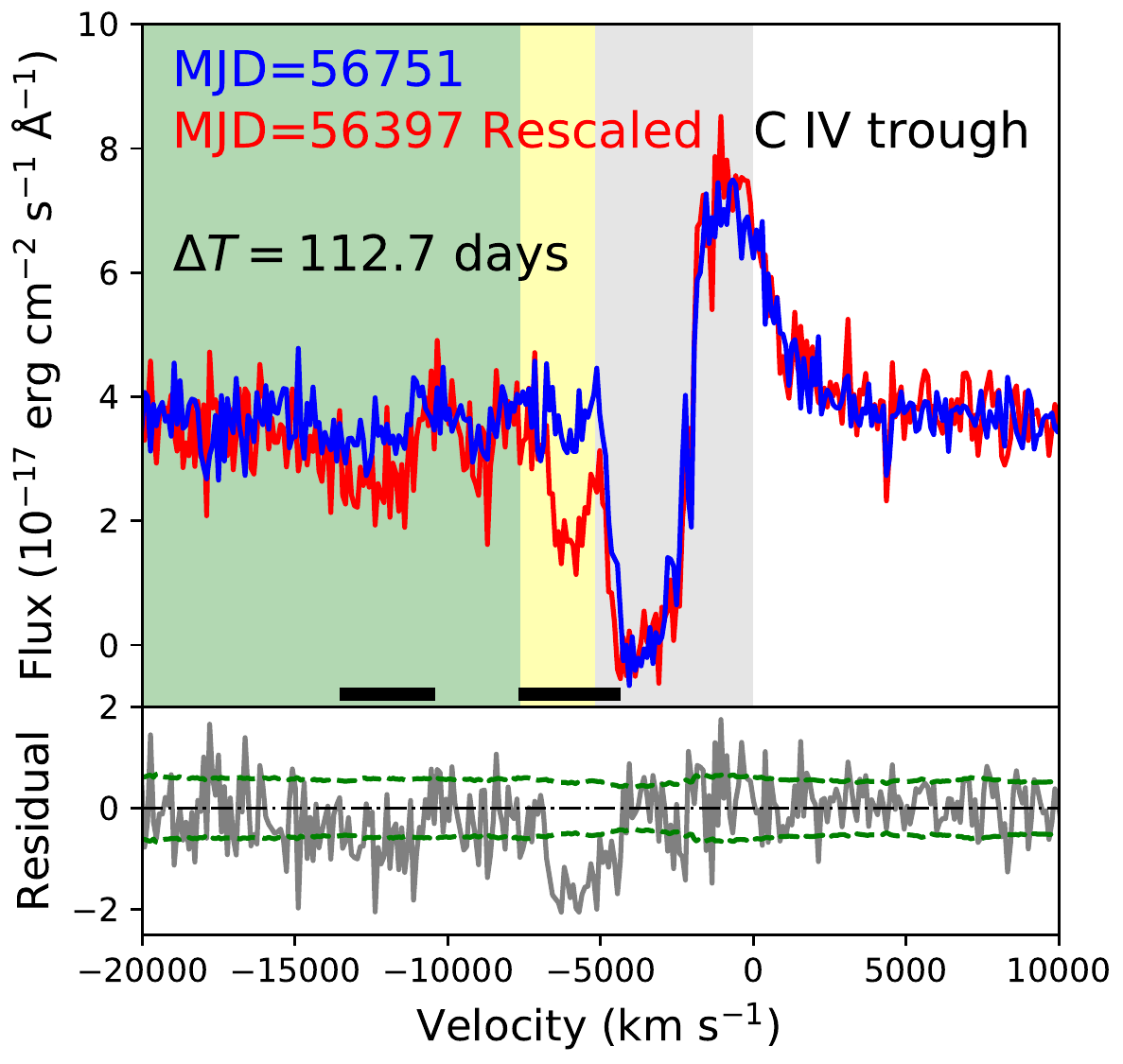}

\caption{Identification of \civ\ variable absorption lines. 
Two examples of matching the reference spectrum (in blue) to another spectrum (in red) by multiplying the reference spectrum with a double power-law described in the text.
The black horizontal lines represent the varied region of \civ\ BAL. 
Left panel is an example for the time interval $\Delta T$= 4.1 days between two observations shorter than 
the recombination timescale $t_{\rm rec}$ of the low- and medium-velocity regions. 
Right panel is an example for the $\Delta T$=112.7 days longer than the $t_{\rm rec}$.
\label{fig3}}
\end{figure*}

\begin{figure}[htb]
\centering
\includegraphics[width=0.45\textwidth]{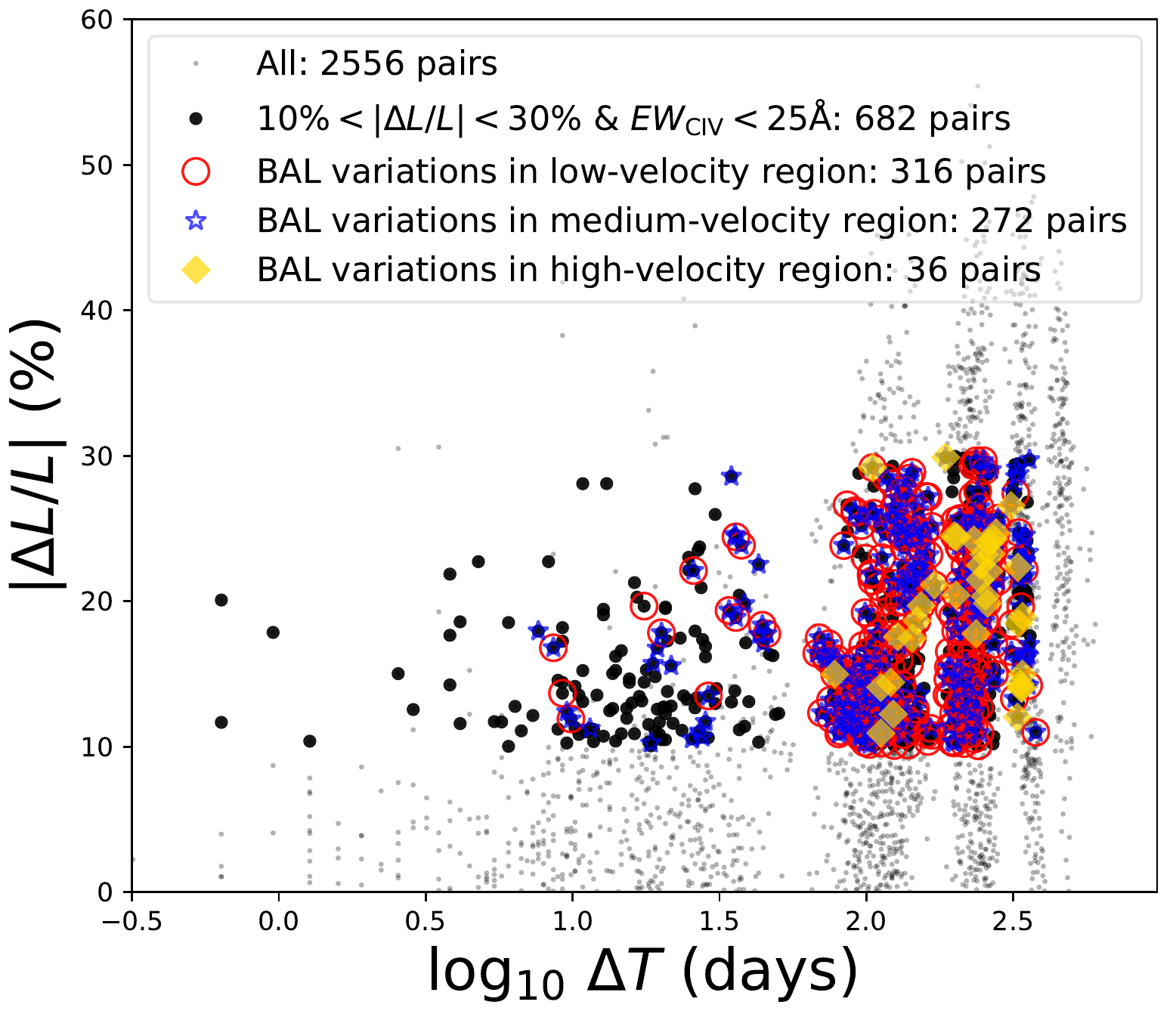}
\caption{The data selection. The 72 observations of SDSS J141955.26+522741.1 yield 2556 spectral pairs (gray points). 
To ensure the amplitudes of continuum variations are similar in different time intervals, we cut the $|\Delta L/L|$ between 10\% and 30\%. 
To reject the deep absorption, we select the mean value of \civ\ BAL EW in each pair of spectra to smaller than 25\AA.
After the above screening, there are 682 pairs of spectra retained (black points). 
In the 682 pairs of spectra, the number of pairs detected to have BAL variations are 316, 272 and 36 for the low- (red circles)
, medium- (blue stars) and high-velocity (yellow squares) regions, respectively.
\label{fig4}}
\end{figure}

\subsection{Identification of the variable regions in BAL troughs}  \label{sec:region}
We follow the same method as \cite{he2019} to identify the variable region of the BAL trough. 
As described in Methods part of \cite{he2019}, we take two main steps to identify the variation region of BAL troughs between a pair of spectra.
(1), we select the higher S/N spectrum of the pair of spectra as a template to match the other spectra
by rescaling it using the double power-law function (Eq. 1 in \citealt{wang2015}) to account for the potential variations of the continuum shape.
Then, we add/subtract a Gaussian to/from the rescaled spectrum to account for variations of the emission line.
As shown in Figure \ref{fig3}, the rescaled template matching produces a better fit outside the absorption line region. 
As a result, we measure the absorption line variability from the difference spectrum.
(2), we search the difference spectrum for the contiguous negative and positive pixels and mark all pixels where the difference 
is greater than $3\sigma$. Adjacent marked pixels are then connected to form a variable region. 
Then we expand such regions into neighboring pixels that have the same sign in the difference spectrum but lie at the less than $3\sigma$ significant level. 
Finally, we merge the neighboring regions with the same variable sign and with a separation of less than four pixels (about 1.5\AA, corresponding to 300 \kms).
Due to the low S/N in the \siiv\ region, we only carry on the identification for \civ\ BALs.
As shown in Figure \ref{fig4}, the 72 observations of SDSS J141955.26+522741.1 yield 2556 spectral pairs. 
The amplitude of continuum variation at 1500\AA\ between a pair of observations
is defined as follows:
\begin{equation}
\frac{\Delta L}{L}=2\frac{L_{2}-L_{1}}{L_{2}+L_{1}},
\end{equation}
where the $L_{1}$ and $L_{2}$ are the flux of the pair of observations.
The amplitudes of continuum variations $|\Delta L/L|$ can affect the BAL variability. So, we cut the $|\Delta L/L|$ between 10\% and 30\% to ensure the 
$|\Delta L/L|$ are similar in different time intervals. The variation of BAL is difficult to detect when the BAL trough is deep (also a low S/N of flux) or saturated. 
So, we select the mean value of \civ\ BAL EW in each pair of spectra to smaller than 25\AA\ to reject the deep absorption.
This action also reduces the difference of BAL EW between different time intervals.
As shown in Figure \ref{fig4}, after this screening, there are 682 pairs of spectra retained.
In the 682 pairs of spectra, the number of pairs detected to have BAL variations are 316, 272 and 36 for the low-, medium- and 
high-velocity regions, respectively.

\subsection{Measuring the fraction $F(\Delta T)$ curve}  \label{sec:fdeltat}

\begin{figure*}[htb]
\centering
\includegraphics[width=0.95\textwidth]{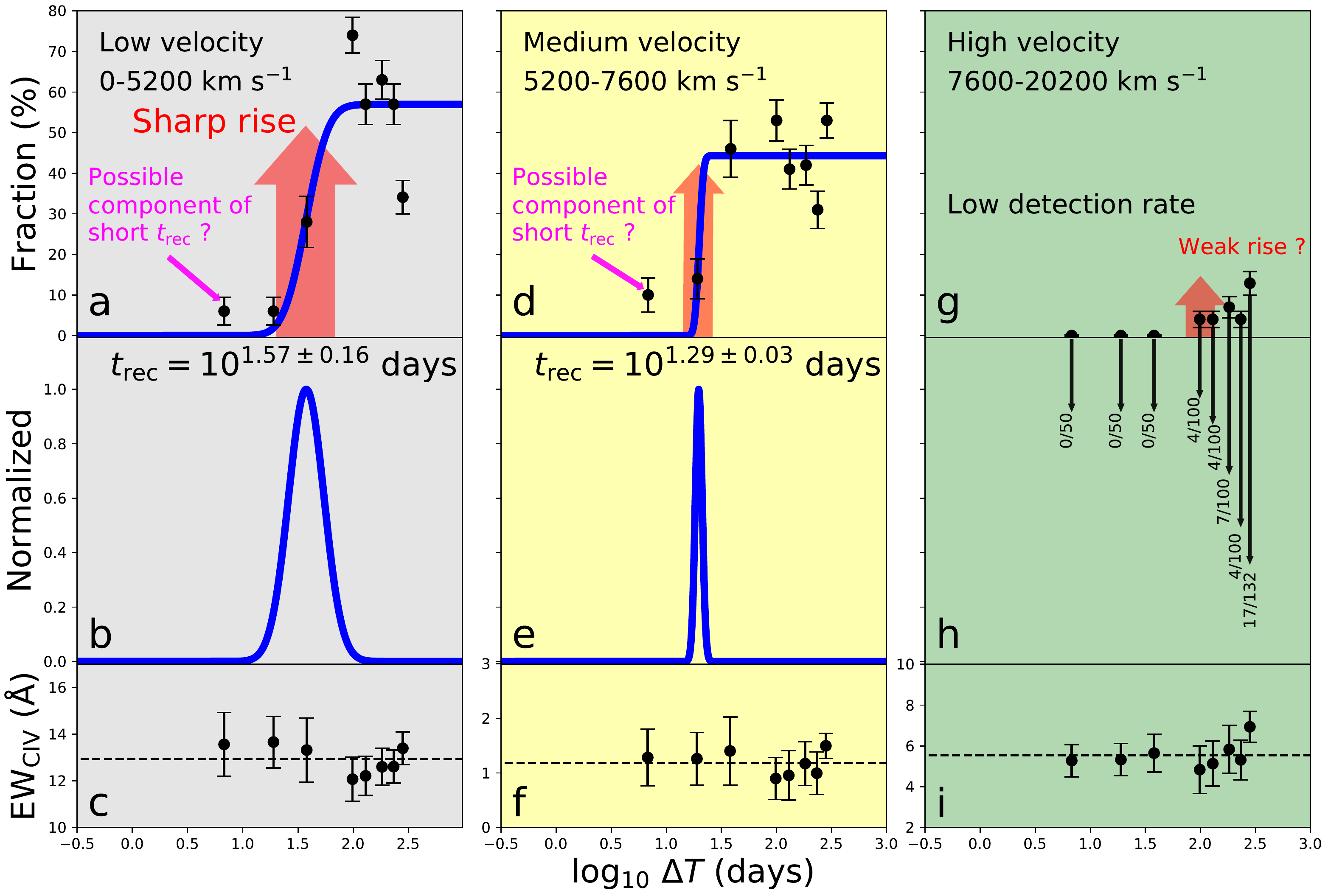}
\caption{The fraction curve of BAL variations of \civ\ in SDSS J141955.26+522741.1.  
For the first time, we detect a "sharp rise" signature in the fraction curve of the low- and medium velocity regions (panel a and d), and
a "weak rise" signature in the high-velocity region  (panel g).
The cumulative distribution function of a Gaussian distribution is used to model the fraction curve. 
Panel b and e: the deduced recombination timescales are $t_{\rm rec}$=10$^{1.57 \pm 0.16}$ and 10$^{1.29 \pm 0.03}$ days for the low- and medium velocity regions, 
respectively. The BAL variations are detected at the time-intervals less than the $t_{\rm rec}$ for half an order of magnitude. 
This result indicates that there may be another component with a shorter recombination timescale in this two individual troughs.
Panel c, f and i: the mean and standard deviation of BAL EWs for all the spectral pairs in each bin. The horizontal dashed line is the 
mean value of BAL EWs for all spectral pairs.
Due to the deep absorption of the low-velocity region, the detection rate decreases as the BAL EW increases in the last five bins.
The absorptions in the medium- and high velocity regions are not deep or saturated. 
As a result, the detection rate does not decrease as the BAL EW increases. }

\label{fig5}
\end{figure*}

To measure the $F(\Delta T)$, we sorted these 682 pairs by the rest time interval between each pair of observations and divided among 8 bins.
Each of the first three bins has 50 spectral pairs. Each of the next four bins has 100 spectral pairs. The last interval has 132 spectral pairs.
For the $i$th bin, the fraction $F_i(\Delta T_i)$ is measured to be $F_i = k_i/N_i$, where $\Delta T_i$ is the mean time interval of spectral pairs,
$k_i$ is the number of spectral pairs with variable BAL, $N_i$ is the number of all spectral pairs in the $i$th bin. 
Since the estimation of $F_i$ amounts to a $N_i$-fold Bernoulli trial, one can use $\sigma_{F_i}=\sqrt{ F_i(1-F_i)/N_i }$
as an estimate of the measurement error of $F_i$. The measured fraction curve is shown in Figure \ref{fig5}. 
As predicted, there is an obvious "sharp rise" signature in the fraction curve for the low- and medium velocity regions of the \civ\ BAL.
In addition, there is a "weak rise" signature in the fraction curve of the high velocity region.
The fraction curve shows two phases. The value of $F(\Delta T)$ at $\log_{10}\Delta T < 1.5$ is significantly lower that at $\log_{10}\Delta T >1.5$
in the fraction curve for the low- and medium velocity regions.

Assume that the error of $t_{\rm rec}$ is a Gaussian function, the fraction $F(\Delta T)$ curve is the cumulative distribution function (CDF) of 
the Gaussian distribution:
\begin{equation}
 F(t)=p_{0}[1+\rm erf(\frac{t-p_{1}}{\sqrt{2}p_{2}})],
 \end{equation}
where $t\equiv\log_{10}\Delta T$ is the logarithmic time interval of spectral pair, 
${\rm erf}(t)=1/\sqrt \pi \int_{-t}^{t}e^{-x^2} dx$ is the error function, and~$p_1=t_{c}$, $p_2=t_{\sigma}$, i.e.,
the mean and standard deviation of the Gaussian distribution, respectively. The best-fit recombination timescale of \civ\ is
$t_{\rm rec}$=10$^{1.57 \pm 0.16}$ and 10$^{1.29 \pm 0.03}$ days for the low- and medium velocity regions, respectively.

As shown bottom panels of in Figure \ref{fig5}, the BAL EWs gradually increase in the last five bins.
Due to the deep absorption of the low-velocity region, the detection rate decreases as the 
BAL EW increases. The absorptions in the medium- and high velocity regions are not deep or saturated. 
As a result, the detection rate does not decrease as the BAL EW increases (shown in panel d and g).

\subsection{Multiple components with different recombination timescales in a single trough?} 
\label{sec:discussion}
As shown in Figure \ref{fig3} and \ref{fig4}, the BAL variations in the low- and medium-velocity regions can be detected at $\Delta T < 10^{1.0}$ days.
Further more, as shown in panel a and d of Figure \ref{fig5}, the fraction $F(\Delta T)$ of the low- and medium-velocity regions is 6\% (3/50) and 10\% (5/50) 
at $\Delta T =10^{0.8}$ days, respectively. However, the deduced recombination timescales are $t_{\rm rec}$=10$^{1.57 \pm 0.16}$ and 10$^{1.29 \pm 0.03}$ 
days for the low- and medium velocity regions, respectively.
As a result, the BAL variations are detected at the time-intervals less than the $t_{\rm rec}$ for half an order of magnitude.
This result indicates that there may be another component with a shorter recombination timescale in this two individual troughs. 
More observations at shorter time-intervals are needed to determine at what the time scale the detection rate falls to near zero.

In addition, the low detection rate of BAL variation prevents us from extracting the recombination timescale in the high-velocity region.
There is only a "weak rise" between $\Delta T =10^{1.5}$ and $10^{2.5}$ days in the high-velocity region.
More observation with longer intervals are also needed to improve the statistical significance of the "weak rise" signature.

\section{Conclusions} 
\label{sec:conclusions}

In this work, we analyse the BAL variations in SDSS J141955.26+522741.1, a BAL quasar at $z=2.145$ with 72 observations from
the SDSS DR16. The strong correlations between the BAL troughs and the continuum allow us to analyze the 
photoionization-driven BAL variations. Our results can be summarized as follows:

\begin{itemize}

\item[1.] As predicted by \citep{Barlow1992,Krolik1995,wang2015, he2019}, the detection rate of BAL variations is a quasi step function.
For the first time, we detect such an obvious "sharp rise" signature in the fraction curve of \civ\ BAL variations of two individual velocity 
components. The recombination timescale of \civ\ deduced from the fraction curve is $t_{\rm rec}$=10$^{1.57 \pm 0.16}$ and 
10$^{1.29 \pm 0.03}$ days for the low- and medium velocity regions, respectively. In addition,
there is a "weak rise" between $\Delta T =10^{1.5}$ and $10^{2.5}$ days the high-velocity region.
More observation with longer intervals are needed to improve the statistical significance of the "weak rise" signature.

\item[2.] In the low- and medium-velocity troughs, the BAL variations are detected at the time-intervals less than the $t_{\rm rec}$ for half 
an order of magnitude. This result indicates that there may be another component with a shorter recombination timescale in the 
individual trough. However, the statistical significance of the current data is insufficient to confirm this conclusion. 
In the future, more observations at short time-intervals may reveal the answer.

\end{itemize}

\acknowledgments
We thank the anonymous referee for the valuable comments and constructive suggestions.
Z.-C. H. is supported by NSFC-11903031 and USTC Research Funds of the Double First-Class Initiative YD 3440002001.
G.-L. L. acknowledges the grant from the National Natural Science Foundation of China (No. 11673020 and No. 11421303) and 
the Ministry of Science and Technology of China (National Key Program for Science and Technology Research and Development, 
No. 2016YFA0400700). H. -X. G. acknowledges the NSF grant AST-1907290.
G.M. acknowledges “the Fundamental Research Funds for the Central Universities” (No. 2042019kf0040).
This work is also supported by NSFC (11833007, 11733001, 11703022, 11973002, 11873045, 11822301).

Funding for the Sloan Digital Sky 
Survey IV has been provided by the 
Alfred P. Sloan Foundation, the U.S. 
Department of Energy Office of 
Science, and the Participating 
Institutions. 

SDSS-IV acknowledges support and 
resources from the Center for High 
Performance Computing  at the 
University of Utah. The SDSS 
website is www.sdss.org.

SDSS-IV is managed by the 
Astrophysical Research Consortium 
for the Participating Institutions 
of the SDSS Collaboration including 
the Brazilian Participation Group, 
the Carnegie Institution for Science, 
Carnegie Mellon University, Center for 
Astrophysics | Harvard \& 
Smithsonian, the Chilean Participation 
Group, the French Participation Group, 
Instituto de Astrof\'isica de 
Canarias, The Johns Hopkins 
University, Kavli Institute for the 
Physics and Mathematics of the 
Universe (IPMU) / University of 
Tokyo, the Korean Participation Group, 
Lawrence Berkeley National Laboratory, 
Leibniz Institut f\"ur Astrophysik 
Potsdam (AIP),  Max-Planck-Institut 
f\"ur Astronomie (MPIA Heidelberg), 
Max-Planck-Institut f\"ur 
Astrophysik (MPA Garching), 
Max-Planck-Institut f\"ur 
Extraterrestrische Physik (MPE), 
National Astronomical Observatories of 
China, New Mexico State University, 
New York University, University of 
Notre Dame, Observat\'ario 
Nacional / MCTI, The Ohio State 
University, Pennsylvania State 
University, Shanghai 
Astronomical Observatory, United 
Kingdom Participation Group, 
Universidad Nacional Aut\'onoma 
de M\'exico, University of Arizona, 
University of Colorado Boulder, 
University of Oxford, University of 
Portsmouth, University of Utah, 
University of Virginia, University 
of Washington, University of 
Wisconsin, Vanderbilt University, 
and Yale University.



\bibliography{./ref.bib}


\end{document}